\documentclass{PoS}
\usepackage{bm}
\usepackage{graphicx}
\usepackage{caption}
\usepackage{subcaption}

\def\IM{{\rm Im}}

\def\kv{{\bf{k}}}

\def\xu{\hat{\bf{x}}}
\def\yu{\hat{\bf{y}}}
\def\zu{\hat{\bf{z}}}

\def\barr{\begin{eqnarray}}
\def\earr{\end{eqnarray}}
\def\kpt{  \kv'_{\rm T} }

\def\kptkpt{{\kv'}^2_{\rm T}}

\def\bt {b\T}
\def\bl {b\L}
\def\T{_{\rm T}}
\def\L{_{\rm L}}

\def\be{\begin{equation}}
\def\ee{\end{equation}}

\title{Inclusion of the $^3P_0$ model in PYTHIA 8}

\ShortTitle{Inclusion of the $^3P_0$ model in PYTHIA 8}

\author{Albi Kerbizi\thanks{Speaker.}\\
        \rm{University of Trieste and INFN}\\
        E-mail: \email{albi.kerbizi@ts.infn.it}}

\author{Leif L\"onnblad\\
        \rm{Lund University}\\
        E-mail: \email{leif.lonnblad@thep.lu.se}}

\abstract{The spin effects in the hadronization process have been included for the first time in the PYTHIA 8 event generator. The spin effects are limited to the production of pseudo-scalar mesons and are obtained from the propagation of the quark polarization along the fragmentation chain according to the rules of the so-called $^3P_0$ model. The interface between PYTHIA 8 and the package of the $^3P_0$ model is presented together with preliminary results on the Collins and di-hadron asymmetries as obtained from simulations of the transversely polarized semi-inclusive deep inelastic scattering process.}

\FullConference{The XXVII International Workshop on Deep Inelastic Scattering and Related Subjects (DIS2019)\\
                8-12 April, 2019\\
                Torino, Italy}

\begin{document}

\section{Introduction}
Experiments dedicated to the study of the nucleon spin structure have recently shown that the quark spin is at the origin of important observable effects.
In particular, quarks inside a transversely polarized nucleon can have a transverse polarization which is described by the chiral odd transversity parton distribution function (PDF) $h_1^q$, the less known among the three PDFs necessary for the description of the collinear nucleon structure at leading twist.
In addition the hadrons produced in the fragmentation process of a transversely polarized quark show a left-right asymmetry with respect to the plane defined by the quark momentum and its polarization vector. This phenomenon is known as the Collins effect and it is described by the chiral odd Collins fragmentation function (FF) $H_{1q}^{\perp h}$ \cite{collins}.
In a transversely polarized semi-inclusive deep inelastic scattering (SIDIS) process these two effects give rise to a specific modulation in the azimuthal distribution of the final hadrons, which amplitude is known as the Collins asymmetry.
The Collins asymmetry gives access to $h_1^q$, provided that $H_{1q}^{\perp h}$ is known. It is measured in $e^+e^-$ annihilation. Since the Collins function is a non perturbative object, it can be tackled through models and Monte Carlo simulations. However, presently, these effects are not taken into account in the most popular event generators.

Recently a $^3P_0$ model of polarized quark fragmentation \cite{kerbizi-2018,kerbizi-2019} has been implemented as a stand alone Monte Carlo code. The model is based on the Lund Model of string hadronization \cite{andersson-1983} already implemented in PYTHIA, with an additional assumption that the $q\bar{q}$ pairs at string breakups are produced with the vacuum quantum numbers $L=0$, $S=1$ and $J=0$. In addition to the phenomenological parameters of the Lund Model, the $^3P_0$ model introduces only one further complex parameter responsible for the Collins and other spin effects. Presently, only the production of pseudo-scalar mesons is considered.

The stand alone Monte Carlo \cite{kerbizi-2019} has been integrated in PYTHIA 8 \cite{pythia8}, allowing, for the first time, to introduce spin effects in this event generator.
The interface between the $^3P_0$ model and PYTHIA, the validation of the implementation of the spin effects in PYTHIA and the results of the first simulation of the Collins and di-hadron asymmetries are presented. The description of the $^3P_0$ model can be found in Ref. \cite{kerbizi-2019}.

\section{The interface between PYTHIA and the $^3P_0$ model}\label{sec:interface}

We let first PYTHIA generate the hard scattering in the $\gamma^*$-nucleon (GNS) reference system, namely in the system where the exchanged $\gamma^*$ propagates along $\zu$ axis and the target is at rest. The initial and final beam momenta lie in the $(\xu,\zu)$ plane defined by the lepton scattering plane. An active quark with flavour $q_{act}$ is picked up from the target and after the hard scattering PYTHIA sets up a string between this quark and the remnant system, for instance a di-quark $qq$. In the GNS we define also the polarization vector $\textbf{S}_{q}$ of the active quark and the polarization vector $\textbf{S}_{q}'$ of $q_{act}$ after hard scattering. The polarization of the target remnant is neglected. The polarization vector and the string endpoints are then boosted to the string rest system, namely the reference system where the momenta of the endpoints are opposite and directed along the $\zu$ axis. The string is now ready to hadronize and we force hadronization to evolve from the quark side towards the remnant side.

PYTHIA starts the hadronization by generating a first break-up with a $q'\bar{q}'$ pair, and then forms the hadron $h(q_{act}\bar{q}')$ with momentum $p$. We accept the hadron according to the weight
\begin{eqnarray}\label{eq:w}
    w(\kpt,\textbf{S}'_q)=\frac{1}{2}\left(1-\frac{2\IM(\mu)\textbf{S}'_q\cdot(\zu\times\kpt)}{|\mu|^2+\kptkpt} \right),
\end{eqnarray}
obtained from the splitting function of the $^3P_0$ model in Ref. \cite{kerbizi-2019}. This weight can be thought as the probability of having a $^3P_0$ break-up. It depends on the complex mass parameter $\mu$ of the $^3P_0$ model \cite{kerbizi-2019}, on the transverse momentum $\kpt$ of $q'$ and on the polarization vector $\textbf{S}'_q$. The current hadron is rejected with probability $1-w$ and if it is not pseudo-scalar. When rejected a new one is tested until success. The accepted hadron is then stored in the event record and the polarization vector of $q'$ is calculated using the rules of the $^3P_0$ model \cite{kerbizi-2019}. The chain of break-ups is continued in this way until the energy left in the remaining string piece is sufficient only for the production of the last two hadrons that complete the hadronization of the initial string. This exit condition is handled by the standard procedure of PYTHIA without any external intervention. The last two hadrons are therefore produced without spin effects.

\section{Simulation of transversely polarized DIS events}\label{sec:polDIS}

In this section we show results obtained from simulations of DIS events where muons with $160\,\,\rm{GeV}/c$ momentum are scattered off a transversely polarized proton (or neutron) target. The phase space cuts on the usual DIS variables $W>5\,\,\rm{GeV}$, $0.2<y<0.9$ and $Q^2>1.0\,\,\rm{GeV^2}$ have been applied. This kinematic regime is that of the COMPASS experiment \cite{compass_collins_pi_k_proton}. The PYTHIA parameters $\rm{StringZ:aLund}$, $\rm{StringZ:bLund}$, $\rm{StringPT:sigma}$ have been set to the values $a=0.9$, $\bl=0.5\,GeV^{-2}$ and $\bt^{-1/2}=0.34\,GeV$ used in the stand alone $^3P_0$ Monte Carlo program \cite{kerbizi-2019}. The complex mass has been taken $\mu=(0.42+\rm{i}0.76)\,\,GeV$ \cite{kerbizi-2019}.
In the analysis we apply also cuts on the hadron fractional energy $z_h>0.2$ and transverse momentum $\rm{p_T}>0.1\,\rm{GeV}/c$ in the GNS. The $z_h$ cut is lowered to $z_h>0.1$ when we look at pairs of hadrons in the same jet.
For the simulations performed here, we have also neglected strings which fragment in three hadrons where only one is weighted according to the $^3P_0$ mechanism and the other two are produced by the exit procedure. 

To ensure that the spin effects are introduced correctly in PYTHIA, we compare the relevant observables as generated with the stand alone $^3P_0$ MC and with PYTHIA interfaced to the $^3P_0$ model (PYTHIA+$^3P_0$).
With the stand alone MC we generate only jets of fully transversely polarized $u$ quarks whereas in PYTHIA we select only DIS events where the string is stretched between a fully transversely polarized $u$ quark and a scalar diquark $(ud)_0$.

\begin{figure}[h]
        \centering
        \begin{minipage}{.75\textwidth}
        \centering
        \includegraphics[width=0.8\textwidth]{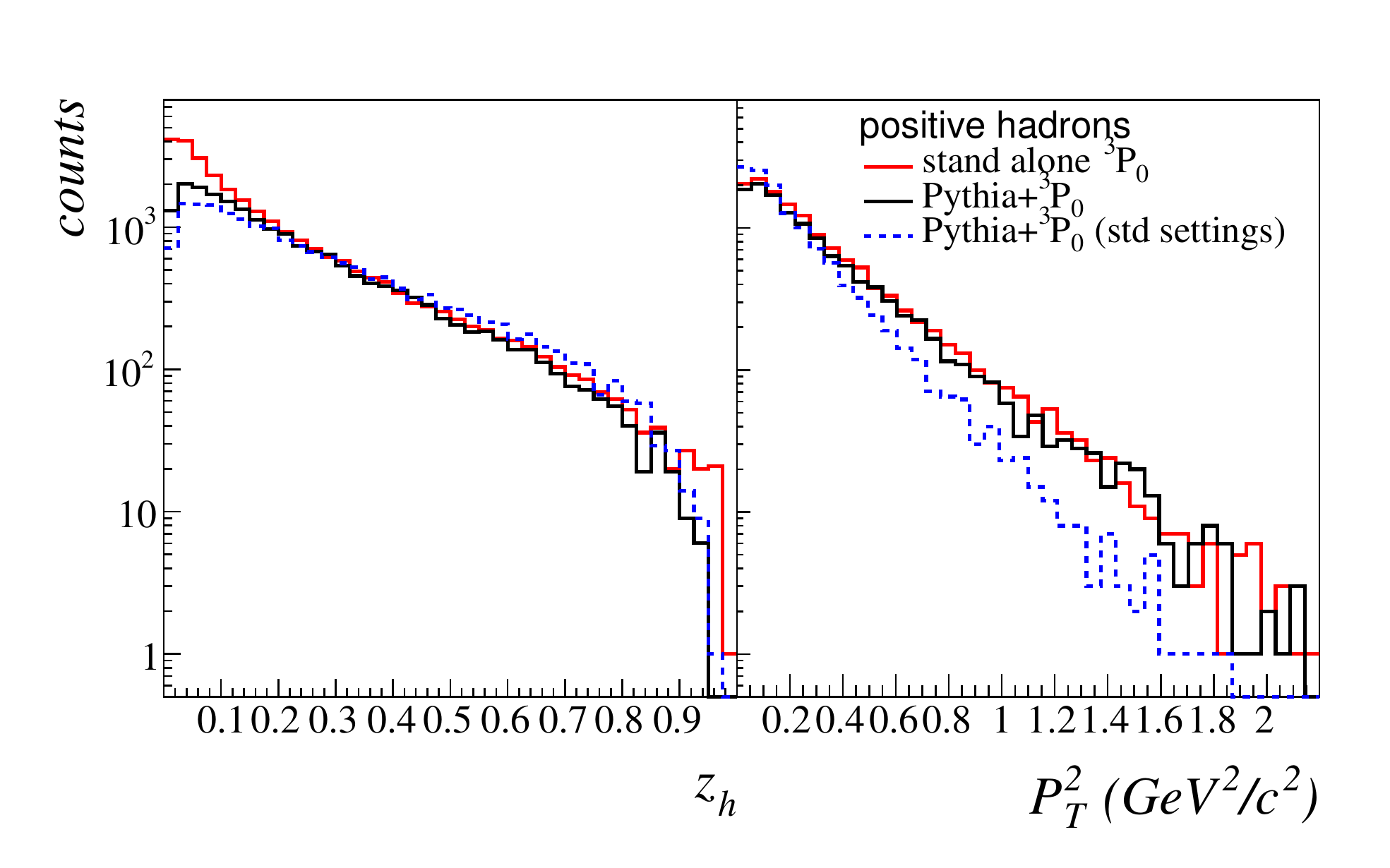}
        \end{minipage}%
        \caption{Distributions of the hadrons fractional energy $z_h$ (left plot) and transverse momentum (right plot) as obtained with the stand alone $^3P_0$ MC (solid red histogram), with PYTHIA interfaced with the $^3P_0$ model with $^3P_0$ settings (solid black histogram) and with standard settings (dotted blue histogram).}
        \label{fig:validation}
\end{figure}

In Fig. \ref{fig:validation} we compare the $z_h$ (left plot) and $\rm{p^2_T}$ (right plot) distributions as obtained with the stand alone $^3P_0$ MC (solid red histogram) and with PYTHIA+$^3P_0$ (solid black histogram) for positive hadrons. Some differences can be seen for small $z_h$, due to the different exit conditions in the two MC programs, and for large $z_h$, due to the presence of the heavy $(ud)_0$ in PYTHIA. These distributions are also very similar to those obtained with PYTHIA+$^3P_0$ but with standard PYTHIA settings (dotted blue histogram). Only a slight difference can be seen in the $\rm{p^2_T}$ distribution. Also we checked that the interface with the $^3P_0$ model does not affect the $z_h$ and the $\rm{p}^2\T$ distributions of standard PYTHIA.

The Collins analysing power is calculated as $a^{u\uparrow \rightarrow h +X}=2\langle\sin\phi_C\rangle$, where $\phi_C=\phi_h-\phi_{\textbf{S}_{q}}$ is the Collins azimuthal angle, as function of $z_h$ and $\rm{p_T}$. The two simulations give very much the same results. The results on the di-hadron asymmetries are also the same, meaning that the spin effects of the $^3P_0$ model are correctly implemented in PYTHIA.

The transverse polarization of the active quark before hard scattering is $\textbf{\rm{S}}_{\rm{qT}} = \frac{h_1^q}{f_1^q}\,\rm{\textbf{S}_{NT}}$,
where $\rm{\textbf{S}_{NT}}$ is the transverse polarization vector of the target nucleon in the GNS, $h_1^q$ is the transversity PDF and $f_1^q$ is the unpolarized quark PDF.
After the interaction, the polarization vector of the scattered quark is reflected with respect to the normal to the lepton scattering plane, which in the GNS system coincides with $\yu$ axis. The transverse polarization is also decreased by the depolarization factor $D_{NN}=(1-y)/(1-y+y^2/2)$, $y$ being the fraction of the initial lepton energy carried by the exchanged virtual photon.
This is simulated by setting the polarization vector of the fragmenting quark to be
\begin{equation}\label{eq:Sq_scattered}
    \textbf{\rm{S}}'_{\rm{qT}} = D_{NN}\,\left[\textbf{\rm{S}}_{\rm{qT}}-2(\textbf{\rm{S}}_{\rm{qT}}\cdot\xu)\xu\right].
\end{equation}
The transverse polarization can be implemented for each quark and anti-quark and the parametrizations of $h_1^q$ and $f_1^q$ can be chosen by the user.
Presently the transverse polarization is implemented for $q=u_v,d_v$ using the parametrizations $xh_1^{u_v} = 3.2\,x^{1.28}\,(1-x)^4$ and $xh_1^{d_v}=-4.6\,x^{1.44}\,(1-x)^4$ \cite{martin-barone-bradamante}.

For the unpolarized PDF $f_1^q$ the parametrization of CTEQ5L LO which is set as default in PYTHIA 8 has been used.

\begin{figure}
\centering
\begin{minipage}{.95\textwidth}
  \centering
  \includegraphics[width=.8\linewidth]{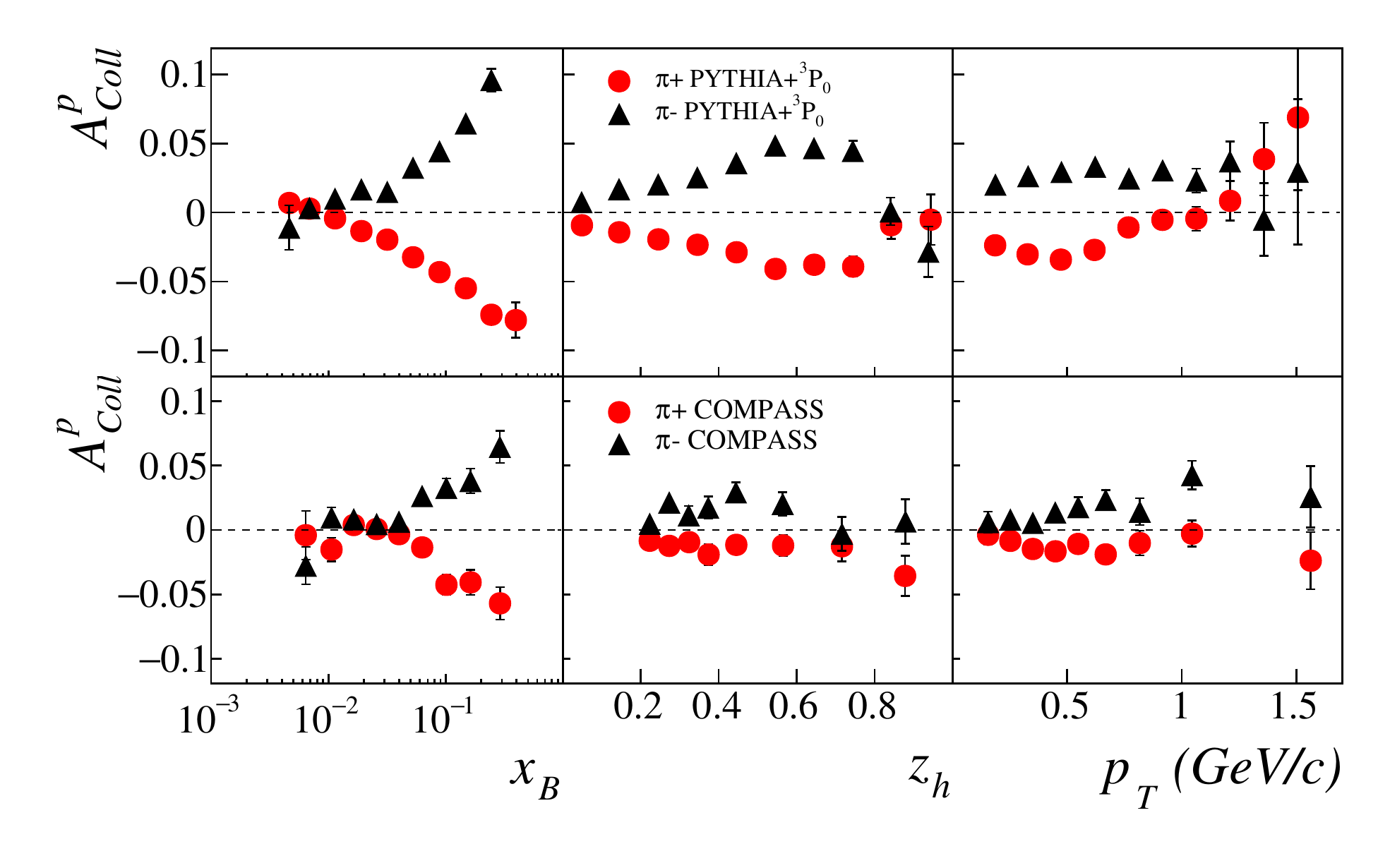}
\end{minipage}%
\caption{Collins proton asymmetry as function of $x_B$, $z_h$ and $\rm{p_T}$ for positive pions (circles) and negative pions (triangles) as obtained from PYTHIA+$^3P_0$ (upper row) compared to COMPASS data \cite{compass_collins_pi_k_proton} (lower row).}
\label{fig:collins_asymmetry_p}
\end{figure}

The Collins asymmetry for positive pions (circles) and negative pions (triangles) for a proton target as obtained from simulations with PYTHIA+$^3P_0$ with the parameters of the stand-alone $^3P_0$ Monte Carlo at the COMPASS energy, is shown in the upper row of Fig. \ref{fig:collins_asymmetry_p}. The corresponding COMPASS proton asymmetry \cite{compass_collins_pi_k_proton} is shown in the lower row of the same figure. As can be seen, the trends as function of $x_B$ and $\rm{p_T}$ are similar. Some differences can be seen at intermediate values of $z_h$, where the decay of polarized vector mesons is expected to contribute. Also, the asymmetry from PYTHIA+$^3P_0$ is somewhat larger than data and some retuning, e.g. $\IM(\mu)$, is needed. All in all the comparison with the experimental data is very satisfactory.

The corresponding results have been produced also for a deuteron target by merging simulations performed separately for a proton and for a neutron target.
From simulations we obtain very small values (below $1\%$) for the Collins asymmetry on deuteron in very good agreement with the COMPASS data \cite{compass_collins_pi_k_deuteron}.

\begin{figure}
\centering
\begin{minipage}{.95\textwidth}
  \centering
  \includegraphics[width=.8\linewidth]{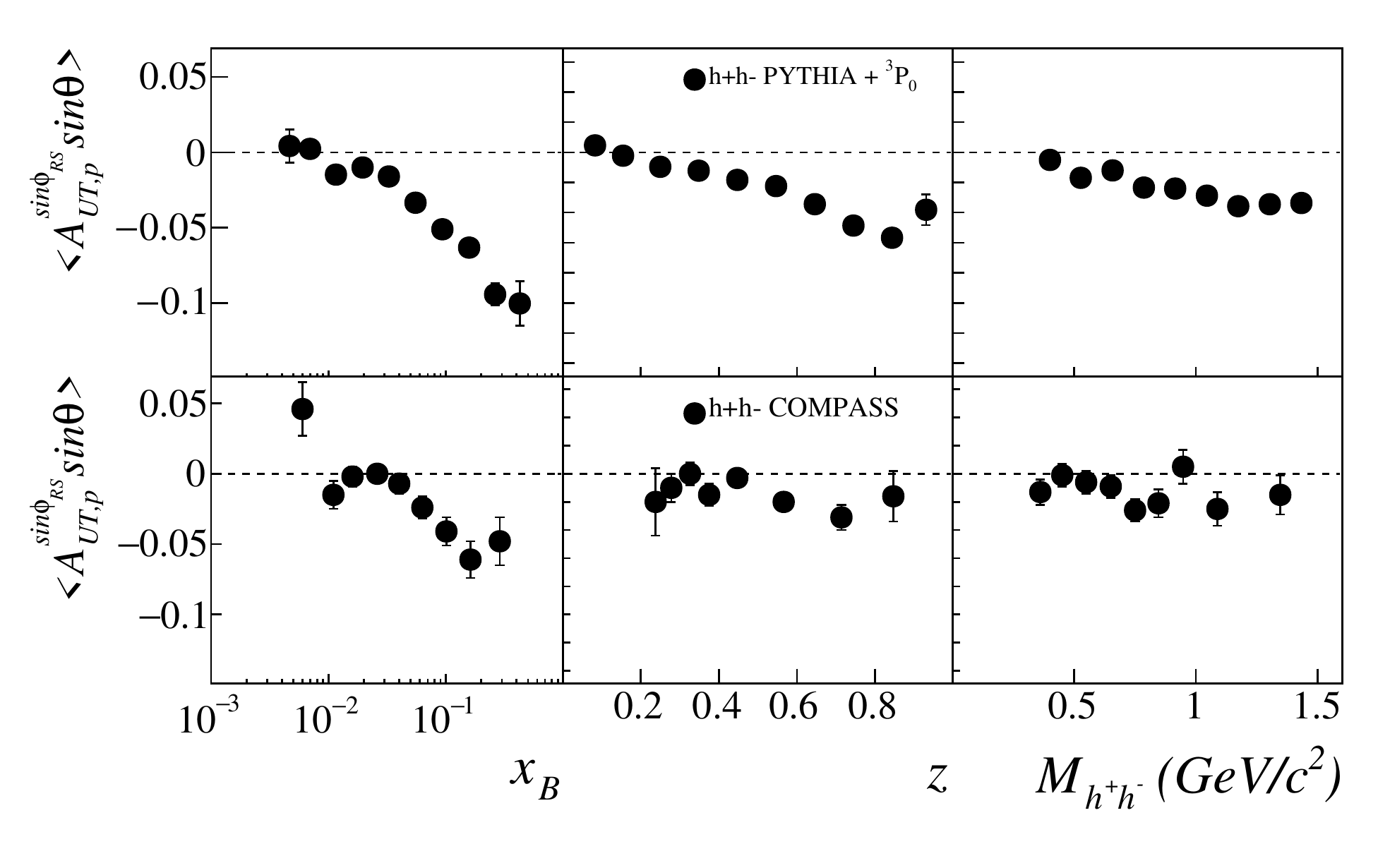}
\end{minipage}%
\caption{Di-hadron asymmetry for pairs of oppositely charged hadrons in the same jet, as function of $x_B$, of the sum of their fractional energies $z$ and of the invariant mass, as obtained with PYTHIA+$^3P_0$ on a proton target (upper row) and the COMPASS \cite{compass_2h_proton} data (lower row).}
\label{fig:2h_asymmetry_p}
\end{figure}

The same simulated data have been used to evaluate the so called di-hadron asymmetry, namely an azimuthal asymmetry defined for oppositely charged hadrons pairs in the same jet. It is shown in Fig. \ref{fig:2h_asymmetry_p} (upper row) for a proton target as function of $x_B$, of $z=z_{h_1}+z_{h_2}$ and of the invariant mass of the pair. The trends are similar as in COMPASS \cite{compass_2h_proton} data (lower row), again with a slightly larger amplitude. The di-hadron asymmetry for a deuteron target is again very small in agreement with the COMPASS data \cite{compass_2h_deuteron}.

\section{Conclusions}
For the first time the quark spin has been implemented in PYTHIA 8. The $^3P_0$ model of polarized quark fragmentation has been used. The transversely polarized SIDIS process where the transverse polarization of the initial quark is calculated introducing a parametrization for the transversity PDF has been simulated. The simulations have been performed at the COMPASS kinematics obtaining Collins and di-hadron asymmetries which compare well with the COMPASS results.
This has to be regarded as a first step towards the account of spin effects in PYTHIA which can be used to perform multi-dimentional studies, compare with experimental data and make predictions for future experiments. A write-up is in preparation and PYTHIA+$^3P_0$ will be available soon.

\section*{Acknowledgments}
We thank X. Artru, F. Bradamante, M. Diefenthaler and A. Martin for useful discussions and suggestions. Part of the work has been supported by the Jefferson Lab project "Phenomenological study of hadronization in nuclear and high-energy physics experiments".

\end{document}